\documentclass[11pt,reqno]{amsart}
\allowdisplaybreaks[4]
\usepackage{graphicx} 
\usepackage{epsfig}
\usepackage{amssymb}
\usepackage{amsmath}
\usepackage{cite}
\newtheorem{theorem}{Theorem}

\newtheorem{proposition}[theorem]{Proposition}
\newtheorem{corollary}[theorem]{Corollary}

\newtheorem{lemma}[theorem]{Lemma}
\newcommand{\be}{\begin{equation}}
\newcommand{\ee}{\end{equation}}
\newcommand{\bea}{\begin{eqnarray}}
\newcommand{\eea}{\end{eqnarray}}
\newcommand{\ba}{\begin{array}}
\newcommand{\ea}{\end{array}}
\newcommand{\bean}{\begin{eqnarray*}}
\newcommand{\eean}{\end{eqnarray*}}

\newcommand{\pa}{\partial}

\begin{document}

\title[Integrability ]{The ``Ghost" Symmetry of the BKP hierarchy}
\author{Jipeng Cheng\dag,Jingsong He$^*$ \ddag, Sen Hu\dag
 }
\dedicatory {  \dag \ Department of Mathematics, USTC, Hefei,Anhui
230026  ,
P.\ R.\ China\\
\ddag \ Department of Mathematics, Ningbo University, Ningbo ,
Zhejiang 315211, P.\ R.\ China }

\thanks{$^*$Corresponding author. email:hejingsong@nbu.edu.cn; jshe@ustc.edu.cn.}
\begin{abstract}
In this paper, we systematically develop the``ghost" symmetry of the
BKP hierarchy through its actions on the Lax operator $L$, the
eigenfunctions and the $\tau$ function. In this process, the
spectral representation of the eigenfunctions and a new potential
are introduced by using squared eigenfunction potential(SEP) of the
BKP hierarchy. Moreover, the bilinear identity of the constrained
BKP hierarchy and Adler-Shiota-van-Moerbeke formula of the BKP
hierarchy are re-derived compactly by means of the spectral
representation and ``ghost" symmetry.

\end{abstract}
\maketitle
\section{Introduction}
Symmetry\cite{olver1} plays an important role in the study of the
integrable system. Many crucial properties of the integrable system,
such as the Noether conserved laws, Hamiltonian structure, Darboux
transformation and reduction, are closely connected with symmetries.
There are several kinds of symmetry of the integrable system. For
instance,in the well-known KP theory, there is an important symmetry
called ``ghost" symmetry \cite{Oevel93}. By identifying the ``ghost"
symmetry with the $k$-th time flow, the constrained KP hierarchy
(cKP)\cite{orlov1993,orlovphysd1993,2OC93,Cheng92,Cheng91,KSS,KS,SS91,SS93}
can be easily defined. In this paper, we shall focus on the study of
the ``ghost" symmetry of the BKP hierarchy.

The ``ghost" symmetry was first introduced by W.Oevel \cite{Oevel93}
in studying the solutions of the cKP hierarchy. Then it was
extensively studied in
\cite{2OC93,OR93,1OC93,OS94,Aratyn98,LorisWillox99}. In the KP
hierarchy, the ``ghost" symmetry is closely related with a squared
eigenfunction potential (SEP),  which is associated to a pair of
arbitrary eigenfunction $\Phi(t)$ and adjoint eigenfunction
$\Psi(t)$ by means of following definition \cite{Oevel93}:
\begin{equation}\label{sepforKP}
\dfrac{\partial}{\partial t_n}S(\Phi(t),\Psi(t))=
Res(\partial^{-1}\Psi M_n \Phi \partial^{-1}).
\end{equation}
Here \footnote{In this paper,we use the notations:
$(\sum_ia_i\pa^i)_+=\sum_{i\geq 0}a_i\pa^i$,
$(\sum_ia_i\pa^i)_-=\sum_{i< 0}a_i\pa^i$,
$(\sum_ia_i\pa^i)_{[k]}=a_k$, $Res(\sum_ia_i\pa^i)=a_{-1}$ and
$(\sum_ia_i\pa^i)^*=\sum_i(-\pa)^ia_i$. }$M_n=L^{n}_{+}$ and $L$ is
a Lax operator of the KP hierarchy.  The predecessor of SEP was in
fact the Cauchy-Baker-Akhiezer kernel introduced in \cite{go1989},
which is an important object for a study of vector fields action on
Riemann surfaces and Virasoro action on tau functions. In
\cite{Aratyn98}, Aratyn \emph{et al} gave a systematic study for the
SEP and the ``ghost" symmetry in KP case. By using SEP as a basic
building block in the definition of the KP hierarchy, they
established a new way to reformulate the theory of the KP hierarchy
called SEP method. The crucial fact of the SEP method is that there
exists a spectral representation for any eigenfunction of the KP
hierarchy with SEP as an spectral density. They also showed that the
``ghost" symmetry \cite{2OC93,1OC93}, which is generated by SEP, has
close relation with the additional symmetries of the KP
hierarchy\cite{orlov1987,OS86,D93,D95,ASM94,ASM95,vM94}.In fact, SEP
can be regarded as a generating function for the additional
symmetries of the KP hierarchy  when both eigenfunction $\Phi(t)$
and adjoint eigenfunction $\Psi(t)$ defining the SEP are
Baker-Akhiezer (BA) functions.

In present work, we would like to consider the ``ghost" symmetry for
the BKP hierarchy. Here BKP hierarchy\cite{DJKM} is an important
reduction of the ordinary KP hierarchy under the constraints on the
Lax operator $L^*=-\partial L \partial^{-1}$. In contrast with the
KP hierarchy, the SEP of the BKP hierarchy can not generate directly
a symmetry flow due to the BKP constraints $L^*=-\partial L
\partial^{-1}$. Thus, we have to find a new potential, which is used
to generate the ``ghost" symmetry  of the BKP hierarchy and is
expected to be expressed by SEP. So this new potential is called the
B-type of the squared eigenfunction potential(BSEP).  Fortunately,
as we shall show, the BSEP was first introduced by
Loris\cite{LorisWillox99} in the study of symmetry reduction of the
BKP hierarchy.

Similar to the case of the KP hierarchy\cite{Aratyn98},  before
giving the ``ghost" symmetry of the BKP hierarchy, we need to study
SEP first. Starting from the BKP bilinear identity, we shall show
that there is also a spectral representations for the eigenfunctions
of the BKP hierarchy, i.e., any eigenfunction of BKP hierarchy can
be represented as a spectral integral over BA wave function with a
spectral density expressed in terms of SEP. Then according to the
differential Fay identity of the BKP hierarchy, we get the
expression of the basic SEP(the one whose defining eigenfunctions
are BA functions). Thus we can give the general expressions of SEP
for the BKP hierarchy with the spectral representation. We then
point out the importance of the spectral representations by showing
that it can in fact provide another definition of the  BKP
hierarchy. In other words, we get an equivalent formulation of BKP
hierarchy. We also call it SEP method for the BKP hierarchy.

Next, after BSEP is systematically studied, we define  the ``ghost"
symmetry flows $\partial_{\alpha}$ for the BKP hierarchy by means of
its action on the Lax operator $L$ and the dressing operator $W$.
Furthermore, actions of $\partial_{\alpha}$ on the eigenfunction
$\Phi$ and $\tau$ function are given by BSEP.

At last,we consider applications for above theory. We shall first
derive the bilinear identities for the cBKP hierarchy
\cite{LorisWillox99,LorisWillox97,Cheng94} with the SEP method. And
then by letting eigenfunctions in the BSEP be BA functions,we get
the relation between the ``ghost" symmetry and the additional
symmetry: in this case, the BSEP becomes a generating function for
the additional symmetries of the BKP hierarchy. With the help of
this fact, we shall give a simple and straightforward proof for the
Adler-Shiota-van Moerbeke formula of the  BKP hierarchy\cite{L95,
L96,Tu07,Tu08}.

This paper is organized in the following way.In section 2,some basic
facts about the BKP hierarchy are reviewed. Then,SEP for the BKP
hierarchy is studied in detail in section 3.  After some interesting
properties of the BSEP studied in section 4, the ``ghost" symmetry
for BKP is showed in section 5. At last, we devote section 6 to two
applications on the spectral representation and  the ``ghost"
symmetry.
\section{BKP Hierarchy}
   Here, we shall review some basic facts about the BKP hierarchy \cite{DJKM}. The BKP hierarchy can be defined
in Lax form as \be \pa_{2n+1}L=[B_{2n+1},L],\quad
B_{2n+1}=(L^{2n+1})_+,\quad n=0,1,2,\cdots,
 \label{laxeq} \ee
 where the Lax operator is given by
 \be L=\pa+u_1\pa^{-1}+u_2\pa^{-2}+\cdots, \label{laxop}
 \ee
with coefficient functions $u_i$ depending on the time variables
$t=(t_1=x,t_3,t_5,\cdots)$ and satisfies the
 BKP constraint
 \be
L^*=-\pa L\pa^{-1}. \label{constr} \ee It can be shown \cite{DJKM}
that the constraint (\ref{constr}) is equivalent to the condition
$(B_{2n+1})_{[0]}=0$ .

The Lax equation (\ref{laxeq}) is equivalent to the compatibility
condition of the linear system \footnote{For a differential operator
$A$ and a function $f$, $A(f)$ denotes the action of A on f.}\be
L(\psi_{BA}(t,\lambda))=\lambda \psi_{BA}(t,\lambda),\qquad
\pa_{2n+1}\psi_{BA}(t,\lambda)=B_{2n+1}(\psi_{BA}(t,\lambda)),
\label{lineq} \ee where $\psi_{BA}(t,\lambda)$ is called BA wave
function. The whole hierarchy can be expressed in terms of a
dressing operator $W$, so that
\[
L=W\pa W^{-1},\qquad W=1+\sum_{j=1}^\infty w_j\pa^{-j},
\]
and the Lax equation is equivalent to the Sato's equation \be
\pa_{2n+1}W=-(L^{2n+1})_-W, \label{satoeq} \ee with constraint \be
W^*\pa W=\pa. \label{constrsato} \ee Let the solutions of the linear
system (\ref{lineq}) be the form \be
\psi_{BA}(t,\lambda)=W(e^{\xi(t,\lambda)})=w(t,\lambda)e^{\xi(t,\lambda)},
\label{wavefun} \ee where $\xi(t,\lambda)=\sum_{i=0}^\infty
t_{2i+1}\lambda^{2i+1}$ and
$w(t,\lambda)=1+w_1/\lambda+w_2/\lambda^2+\cdots$. Then
$\psi_{BA}(t,z)$ is a wave function of the BKP hierarchy if and only
if it satisfies the bilinear identity\cite{DJKM} \be \int
d\lambda\lambda^{-1}\psi_{BA}(t,\lambda)\psi_{BA}(t',-\lambda)=1,\quad
\forall t, t', \label{bileq} \ee where $\int
d\lambda\equiv\oint_\infty\frac{d\lambda}{2\pi i}=
Res_{\lambda=\infty}$ and $t=(t_1=x,t_3,t_5, \cdots)$.

In the BKP hierarchy,if $\Phi$(or $\Psi$) satisfies
\begin{equation}\label{eigen}
    \pa_{2n+1}\Phi=B_{2n+1}(\Phi)
    \big(or ~\pa_{2n+1}\Psi=-B^*_{2n+1}(\Psi)\big),n=0,1,2,\cdots,
\end{equation}
we shall call $\Phi$ (or $\Psi$) eigenfunction (or adjoint
eigenfunction) of the BKP hierarchy. Obviously,
$\psi_{BA}(t,\lambda)$ is also an eigenfunction. The relation
between the eigenfunctions and adjoint eigenfunctions can be seen
from the fact $B^*_{2n+1}\pa=-\pa B_{2n+1}$. This fact implies that
any eigenfunction $\Phi$ gives rise to an adjoint eigenfunction
$\Psi=\Phi_x$. In particular,we have $\psi^*_{BA}(t,\lambda)=
-\lambda^{-1}\psi_{BA}(t,-\lambda)_x$, where
$\psi^*_{BA}(t,\lambda)\equiv W^{*-1}(e^{-\xi(t,\lambda)})$.
Moreover, from the bilinear identity (\ref{bileq}), solutions of the
BKP hierarchy can be characterized by a single function $\tau(t)$
called $\tau$-function such that\cite{DJKM} \be
w(t,\lambda)=\frac{\tau (t-2[\lambda^{-1}])}{\tau(t)}, \label{tau}
\ee where
$[\lambda^{-1}]=(\lambda^{-1},\frac{1}{3}\lambda^{-3},\cdots)$. This
implies that all dynamical variables $\{u_i \}$ in the Lax  operator
$L$ can be expressed by $\tau$-function, which is an essential
character of the KP and BKP hierarchy. Moreover, another important
property of $\tau$ function of the BKP is the following Fay like
identity.

\begin{proposition}\cite{Tu07} \label{Fay}(Fay identity) The tau function of the BKP hierarchy
satisfies: \bea
&&\sum_{(s_1,s_2,s_3)}\frac{(s_1-s_0)(s_1+s_2)(s_1+s_3)}{(s_1+s_0)(s_1-s_2)(s_1-s_3)}
\tau(t+2[s_2]+2[s_3])\tau(t+2[s_0]+2[s_1])\nonumber\\
&&+\frac{(s_0-s_1)(s_0-s_2)(s_0-s_3)}{(s_0+s_1)(s_0+s_2)(s_0+s_3)}
\tau(t+2[s_0]+2[s_1]+2[s_2]+2[s_3])\tau(t)=0, \label{Fayeq} \eea
where $(s_1,s_2,s_3)$ stands for cyclic permutations of $s_1$, $s_2$
and $s_3$.
\end{proposition}
\begin{proposition}\cite{Tu07} \label{dFay}(Differential Fay
identity)For the BKP hierarchy, \bea
&&\left(\frac{1}{s_2^2}-\frac{1}{s_1^2}\right)
\{\tau(t+2[s_1])\tau(t+2[s_2])-\tau(t+2[s_1]+2[s_2])\tau(t)\}\nonumber\\
&&=\left(\frac{1}{s_2}+\frac{1}{s_1}\right)
\{\pa\tau(t+2[s_2])\tau(t+2[s_1])-\pa\tau(t+2[s_1])\tau(t+2[s_2])\}\nonumber\\
&&\quad+\left(\frac{1}{s_2}-\frac{1}{s_1}\right)\{\tau(t+2[s_1]+2[s_2])\pa\tau(t)-
\pa\tau(t+2[s_1]+2[s_2])\tau(t)\}. \label{dFayeq} \eea
\end{proposition}
\noindent Note that these identities are indeed different from the
counterpart of the KP hierarchy because of the
 BKP constraint (\ref{constr}).  In the next context, we shall  show it is for the same reason that the
 SEP of the BKP hierarchy  can not generate directly the symmetry flow.

\section{SEP for the BKP Hierarchy}

As mentioned in Introduction, we hope to get a new potential- BSEP
from the SEP of the BKP hierarchy. So we shall study some
interesting properties of the SEP of the BKP hierarchy in this
section. For any pair of (adjoint) eigenfunctions
$\Phi(t),\Psi(t)$,there exists a function $S(\Phi(t),\Psi(t))$
called SEP, determined by the following equations,
\begin{equation}\label{sep1}
    \frac{\pa}{\pa t_{2n+1}}S\big(\Phi(t),\Psi(t)\big)=Res\big(\pa^{-1}\Psi(L^{2n+1})_+\Phi\pa^{-1}\big),
n=0,1,2,3,\cdots.
\end{equation}

In particular,for $n=0$,we have,
\begin{equation}\label{sep2}
    \pa_xS(\Phi(t),\Psi(t))=\Phi(t)\Psi(t).
\end{equation}
One can see that this definition is the same as the
one\cite{Oevel93} in the KP hierarchy except even number flows are
frozen. There are two properties of SEP for the BKP hierarchy.
\begin{lemma}If $\Phi(t)$ and $\Psi(t)$ are {\rm BKP} eigenfunction and adjoint eigenfunction respectively, then
one has the following relation:
\begin{eqnarray}
  S(\Phi(t),\psi_{BA}(t,-\lambda)_x)=e^{-\xi(t,\lambda)}(\Phi(t)+\mathcal
{O}(\lambda^{-1})), \label{lemma}\\
   S(\psi_{BA}(t,\lambda),\Psi(t))=e^{\xi(t,\lambda)}(\Psi(t)\lambda^{-1}+\mathcal
{O}(\lambda^{-2})). \label{lemmaa}
\end{eqnarray}

\end{lemma}
\textbf{Proof:}We only prove the first identity since the proof of
the second one is similar. Because
$\psi_{BA}(t,-\lambda)_x=e^{-\xi(t,\lambda)}(-\lambda+\mathcal
{O}(1))$ and
\begin{eqnarray*}
  \int e^{-x\lambda}\Phi(t)dx&=& - \int \lambda^{-1}\Phi(t)d e^{-x\lambda} \\
   &=& -\lambda^{-1}e^{-x\lambda}\Phi(t)+ \lambda^{-1}\int
e^{-x\lambda}\Phi_x(t)dx=\cdots=e^{-x\lambda}(-\lambda^{-1}\Phi(t)+\mathcal
{O}(\lambda^{-2})),
\end{eqnarray*}
we find
 \begin{eqnarray*}
   S(\Phi(t),\psi_{BA}(t,-\lambda)_x)&=& \int \Phi(t)\psi_{BA}(t,-\lambda)_xdx=\int \Phi(t)e^{-\xi(t,\lambda)}(-\lambda+\mathcal {O}(1))dx  \\
    &=& -\lambda\int \Phi(t)e^{-\xi(t,\lambda)}dx+e^{-\xi(t,\lambda)}\mathcal {O}(\lambda^{-1})\\
    &=&-\lambda e^{-\xi(t,\lambda)}(-\lambda^{-1}\Phi(t)+\mathcal
{O}(\lambda^{-2}))+e^{-\xi(t,\lambda)}\mathcal {O}(\lambda^{-1})\\
&=&e^{-\xi(t,\lambda)}(\Phi(t)+\mathcal
{O}(\lambda^{-1})).~~~~~~~~~~~~~{\hspace{5cm} \Box}
 \end{eqnarray*}
\begin{lemma}If $\Phi_1$ and $\Phi_2$ are two eigenfunctions of the BKP hierarchy, then
\begin{equation}\label{lemma1}
\Phi_1\Phi_2=S(\Phi_1,\Phi_{2x})+S(\Phi_2,\Phi_{1x}).
\end{equation}
\end{lemma}
\textbf{Proof:}\footnote{Some useful formulas below are needed in
the proof.
\begin{eqnarray}
  Res(A)&=& -Res(A^*) \label{usefulformua1} \\
  a_x &=& \pa a-a\pa \label{usefulformua2}\\
  Res(A\pa^{-1})&=& A_{[0]} \label{usefulformua3}
\end{eqnarray}
where $A$ is a pseudo-differential operator,and $a$ is a function.}
\begin{eqnarray*}
&&\pa_{t_{2n+1}}S(\Phi_2,\Phi_{1x})= Res(\pa^{-1}\Phi_{1x}B_{2n+1}\Phi_2\pa^{-1}) \\
&=&-Res(\pa^{-1}\Phi_2 B_{2n+1}^*\Phi_{1x}\pa^{-1})~~using ~(\ref{usefulformua1}) \\
&=& Res(\pa^{-1}\Phi_2 \pa
B_{2n+1}\pa^{-1}\Phi_{1x}\pa^{-1})~~using~B_{2n+1}^*=-\pa B_{2n+1} \pa^{-1}\\
&=&Res(\pa^{-1}\Phi_2 \pa
B_{2n+1}\Phi_{1}\pa^{-1})-Res(\pa^{-1}\Phi_2 \pa
B_{2n+1}\pa^{-1}\Phi_{1})~~using ~(\ref{usefulformua2})\\
&=&Res(\Phi_2B_{2n+1}\Phi_{1}\pa^{-1})-Res(\pa^{-1}\Phi_{2x}
B_{2n+1}\Phi_{1}\pa^{-1})+Res(\pa^{-1}\Phi_2 B_{2n+1}^*\Phi_{1})\\
&&~~using~(\ref{usefulformua2})~and~B_{2n+1}^*=-\pa B_{2n+1} \pa^{-1}\\
&=&\Phi_2Res(B_{2n+1}\Phi_{1}\pa^{-1})-Res(\pa^{-1}\Phi_{2x}
B_{2n+1}\Phi_{1}\pa^{-1})+Res(B_{2n+1}\Phi_2 \pa^{-1})\Phi_{1} ~~using ~(\ref{usefulformua1})\\
&=&\Phi_2B_{2n+1}(\Phi_{1})-Res(\pa^{-1}\Phi_{2x}
B_{2n+1}\Phi_{1}\pa^{-1})+B_{2n+1}(\Phi_2) \Phi_{1}~~using ~(\ref{usefulformua3})\\
&=&\Phi_2(\pa_{t_{2n+1}}\Phi_{1})-\pa_{t_{2n+1}}S(\Phi_1,\Phi_{2x})+(\pa_{t_{2n+1}}\Phi_2)
\Phi_{1}\\
&=&\pa_{t_{2n+1}}(\Phi_{1}\Phi_{2})-\pa_{t_{2n+1}}S(\Phi_1,\Phi_{2x})~~~~~~~~~~~~~~~~~\Box
\end{eqnarray*}

\begin{proposition}(Spectral representation)\label{propspectralrep}If $\Phi(t)$ is an eigenfunction of the BKP hierarchy ,then
\begin{equation}\label{srbkp4}
    \Phi(t)=\int d\lambda
\lambda^{-1}\psi_{BA}(t,\lambda)S(\Phi(t'),\psi_{BA}(t',-\lambda)_{x'}),
\end{equation}
where the time $t'$ is taken at some arbitrary fixed value. In other
words, $\Phi(t)$ owns a spectral representation in the form of
\begin{equation}\label{srbkp1}
    \Phi(t)=\int d\lambda \lambda^{-1}\varphi(\lambda)\psi_{BA}(t,\lambda),
\end{equation}
with spectral densities given by SEP,that is,
$\varphi(\lambda)=S\big(\Phi(t'),\psi_{BA}(t',-\lambda)_{x'}\big).$
\end{proposition}
\textbf{Proof:} Denote the RHS of (\ref{srbkp4}) by $I(t,t')$.Then
by the BKP bilinear identity (\ref{bileq}), one finds that
$\pa_{t'_m}I(t,t')=0$.Hence $I(t,t')=f(t)$.  By considering
(\ref{lemma}), we have
 $$I(t,t'=t)=\int d\lambda
\lambda^{-1}\psi_{BA}(t,\lambda)e^{-\xi(t,\lambda)}(\Phi(t)
+\mathcal {O}(\lambda^{-1}))=\Phi(t).\Box$$\\
 \textbf{Remark 1:}Here we only give the spectral representation for
 eigenfunctions.As for the adjoint eigenfunctions,the spectral
 representation can be derived similarly by considering (\ref{bileq}) and (\ref{lemmaa}),that is,
 \begin{equation}\label{adjoint}
    \Psi(t)=\int d\lambda
\psi_{BA}^*(t,\lambda)S(\psi_{BA}(t',\lambda),\Psi(t')).
\end{equation}
However,because of the relation between the
 eigenfunctions and adjoint eigenfunctions,we must show that our spectral representations for BKP hierarchy
 are compatible.

 In fact,any adjoint  eigenfunction $\Psi$ for BKP can be written as the derivative of an eigenfunction $\Phi$,
 that is, $\Psi=\Phi_x$. So with the help of (\ref{srbkp4}), (\ref{bileq}),(\ref{lemma1}) and
$\psi^*_{BA}(t,\lambda)=-\lambda^{-1}\psi_{BA}(t,-\lambda)_x$, then
 \begin{eqnarray*}
   \Phi_x(t)&=&\int d\lambda
\lambda^{-1}\psi_{BA}(t,\lambda)_xS(\Phi(t'),\psi_{BA}(t',-\lambda)_{x'}) \\
   &=&\int d\lambda
\lambda^{-1}\psi_{BA}(t,\lambda)_x\psi_{BA}(t',-\lambda)\Phi(t')-\int
d\lambda
\lambda^{-1}\psi_{BA}(t,\lambda)_xS(\psi_{BA}(t',-\lambda),\Phi(t')_{x'})\\
&=&-\int d\lambda
\lambda^{-1}\psi_{BA}(t,\lambda)_xS(\psi_{BA}(t',-\lambda),\Phi(t')_{x'})\\
&=&-\int d\lambda
\lambda^{-1}\psi_{BA}(t,-\lambda)_xS(\psi_{BA}(t',\lambda),\Phi(t')_{x'})~letting~\lambda\rightarrow-\lambda\\
&=&\int d\lambda
\psi_{BA}^*(t,\lambda)S(\psi_{BA}(t',\lambda),\Phi(t')_{x'}) \\
&=&\int d\lambda
\psi_{BA}^*(t,\lambda)S(\psi_{BA}(t',\lambda),\Psi(t'))=\Psi(t).
 \end{eqnarray*}
So our representation is consistent with $\Psi=\Phi_x$, which shows
it is necessary to only study  the
spectral representation of the  eigenfunctions for the BKP hierarchy.\\
\textbf{Remark 2}:Since
$\psi^*_{BA}(t,\lambda)=-\lambda^{-1}\psi_{BA}(t,-\lambda)_x$,so we
can rewrite (\ref{srbkp4}) as
\begin{equation}\label{eigens}
\Phi(t)=-\int d\lambda
\psi_{BA}(t,\lambda)S(\Phi(t'),\psi_{BA}^*(t',\lambda)).
\end{equation}
Our results (\ref{adjoint}) and (\ref{eigens}) can be regarded as a
natural reduction from corresponding ones \cite{Aratyn98} of the
KP hierarchy by considering BKP constraints $L^*=-\partial L \partial^{-1}$ and $\Psi=\Phi_x$.\\
\textbf{Remark 3}:In particular,
\begin{equation}\label{srbkp6}
\psi_{BA}(t,\mu)=\int d\lambda
\lambda^{-1}\psi_{BA}(t,\lambda)S(\psi_{BA}(t',\mu),\psi_{BA}(t',-\lambda)_{x'})
\end{equation}
is given from (\ref{srbkp4}) by setting $\Phi(t)=\psi_{BA}(t,\mu)$.

Now we shall use the above obtained spectral representation to get
general expressions of SEP.Before this we will use the differential
Fay identity (\ref{dFayeq}) to get
$S(\psi_{BA}(t,\mu),\psi_{BA}(t,-\lambda)_x)$, which is a basic and
useful SEP of the BKP hierarchy. According to the proposition
\ref{dFay}, set$ s_1=\lambda^{-1}$ and $s_2=-\mu^{-1}$, we can find
\begin{eqnarray*}
  &&\pa_x(\frac{\tau(t+2[\lambda^{-1}]-2[\mu^{-1}])}{\tau(t)})\\
  &=& \frac{\pa_x\tau(t+2[\lambda^{-1}]-2[\mu^{-1}])\tau(t)-\tau(t+2[\lambda^{-1}]-2[\mu^{-1}])\pa_x\tau(t)}{\tau^2(t)} \\
   &=& \frac{(-\mu+\lambda)}{(-\mu-\lambda)}\frac{(\pa_x\tau(t-2[\mu^{-1}])\tau(t+2[\lambda^{-1}])-\tau(t-2[\mu^{-1}])\pa_x\tau(t+2[\lambda^{-1}]))}{\tau^2(t)} \\
  &-&\frac{(-\mu+\lambda)(\tau(t-2[\mu^{-1}])\tau(t+2[\lambda^{-1}])-\tau(t+2[\lambda^{-1}]-2[\mu^{-1}])\tau(t)}
  {\tau^2(t)}.
\end{eqnarray*}
Taking into account of the following identity,
\begin{eqnarray*}
&&\pa_x(\frac{\tau(t-2[\mu^{-1}])}{\tau(t)})\tau(t+2[\lambda^{-1}])-\tau(t-2[\mu^{-1}])\pa_x(\frac{\tau(t+2[\lambda^{-1}]}{\tau(t)})\\
  &=&\frac{\pa_x\tau(t-2[\mu^{-1}])\tau(t+2[\lambda^{-1}])-\tau(t-2[\mu^{-1}])\pa_x\tau(t+2[\lambda^{-1}])}{\tau^2(t)},
\end{eqnarray*}
then,
\begin{eqnarray}
  &&\pa_x(\frac{\tau(t+2[\lambda^{-1}]-2[\mu^{-1}])}{\tau(t)})\nonumber  \\
   &=& \frac{\mu-\lambda}{\mu+\lambda}\Big( \pa_x(\frac{\tau(t-2[\mu^{-1}])}{\tau(t)})\tau(t+2[\lambda^{-1}])-\tau(t-2[\mu^{-1}])\pa_x(\frac{\tau(t+2[\lambda^{-1}]}{\tau(t)})\Big)
   \nonumber \\
  &+&(\mu-\lambda)\Big(\frac{\tau(t-2[\mu^{-1}])\tau(t+2[\lambda^{-1}])}{\tau^2(t)}-\frac{\tau(t+2[\lambda^{-1}]-2[\mu^{-1}])}{\tau(t)}\Big).
  \label{partialtauid}
\end{eqnarray}
Furthermore, in order to get
$S(\psi_{BA}(t,\mu),\psi_{BA}(t,-\lambda)_x)$, by using
(\ref{wavefun}) and (\ref{tau}), we first calculate
\begin{eqnarray*}
&&\pa_x(\psi_{BA}(t,\mu) \psi_{BA}(t-2[\mu^{-1}],-\lambda))\\
&=& \pa_x(\frac{\lambda  +
\mu}{\mu-\lambda}e^{\xi(t,\mu)-\xi(t,\lambda)}\frac{\tau(t+2[\lambda^{-1}]-2[\mu^{-1}])}{\tau(t)})\\
 &=&(\lambda  + \mu)e^{\xi(t,\mu)-\xi(t,\lambda)}\frac{\tau(t+2[\lambda^{-1}]-2[\mu^{-1}])}{\tau(t)}-\frac{\lambda  +
\mu}{\lambda-\mu}e^{\xi(t,\mu)-\xi(t,\lambda)}\pa_x(\frac{\tau(t+2[\lambda^{-1}]-2[\mu^{-1}])}{\tau(t)}).
\end{eqnarray*}
Note
\begin{equation}\label{translationidentity}
e^{\xi(-2[\mu^{-1}],-\lambda)}=e^{2\big(\frac{\lambda}{\mu}+(\frac{\lambda}{\mu})^3+
(\frac{\lambda}{\mu})^5+\cdots \big )}=e^{\ln(1+
\frac{\lambda}{\mu}) -\ln(1-\frac{\lambda}{\mu})
}=\frac{\mu+\lambda} {\mu-\lambda}
\end{equation}
 is used in the first equality above. Taking (\ref{partialtauid}) into the last term of above formula,
 then
 \begin{eqnarray*}
&&\pa_x(\psi_{BA}(t,\mu) \psi_{BA}(t-2[\mu^{-1}],-\lambda))\\
&=& (\lambda  + \mu)e^{\xi(t,\mu)-\xi(t,\lambda)}\frac{\tau(t+2[\lambda^{-1}]-2[\mu^{-1}])}{\tau(t)}\\
&-&\frac{\lambda  + \mu}{\lambda-\mu}e^{\xi(t,\mu)-\xi(t,\lambda)}
\Big(\frac{\mu-\lambda}{\mu+\lambda}\{ \pa_x(\frac{\tau(t-2[\mu^{-1}])}{\tau(t)})\tau(t+2[\lambda^{-1}])-\tau(t-2[\mu^{-1}])\pa_x(\frac{\tau(t+2[\lambda^{-1}]}{\tau(t)})\}\\
  &+&(\mu-\lambda)(\frac{\tau(t-2[\mu^{-1}])\tau(t+2[\lambda^{-1}])}{\tau^2(t)}-\frac{\tau(t+2[\lambda^{-1}]-2[\mu^{-1}])}{\tau(t)})\Big)\\
  &=&e^{\xi(t,\mu)-\xi(t,\lambda)}\Big( \pa_x(\frac{\tau(t-2[\mu^{-1}])}{\tau(t)})\tau(t+2[\lambda^{-1}])-\tau(t-2[\mu^{-1}])\pa_x(\frac{\tau(t+2[\lambda^{-1}]}{\tau(t)})\\
  &+&(\mu+\lambda)\frac{\tau(t-2[\mu^{-1}])\tau(t+2[\lambda^{-1}])}{\tau^2(t)}\Big)\\
  &=&\pa_x\psi_{BA}(t,\mu)\psi_{BA}(t,-\lambda)-\psi_{BA}(t,\mu)\pa_x\psi_{BA}(t,-\lambda)\\
  &=&\pa_x\big(\psi_{BA}(t,\mu)\psi_{BA}(t,-\lambda)\big)-2\psi_{BA}(t,\mu)\pa_x\psi_{BA}(t,-\lambda).
\end{eqnarray*}
Note the first term cancels the fourth  term of the first equality
above. So we have,
$$\psi_{BA}(t,\mu)\psi_{BA}(t,-\lambda)_x=\frac{1}{2}\pa_x\{(\psi_{BA}(t,-\lambda)-\psi_{BA}(t-2[\mu^{-1}],-\lambda))\psi_{BA}(t,\mu)\},$$
which implies
\begin{equation}\label{sepbiaozhun}
     S\big(\psi_{BA}(t,\mu),\psi_{BA}(t,-\lambda)_x\big)
     =\frac{1}{2}\big(\psi_{BA}(t,-\lambda)-\psi_{BA}(t-2[\mu^{-1}],-\lambda)\big)\psi_{BA}(t,\mu).
\end{equation}

Next, we shall give the expression of another basic SEP-
$S\big(\Phi(t),\psi_{BA}(t,-\lambda)_x\big)$. According to the
spectral representation of $\Phi(t)$ in (\ref{srbkp1}), then
\begin{eqnarray*}
  S(\Phi(t),\psi_{BA}(t,-\lambda)_x) &=& S(\int d\mu \mu^{-1}\varphi(\mu)\psi_{BA}(t,\mu), \psi_{BA}(t,-\lambda)_x)\\
   &=& \int d\mu \mu^{-1}\varphi(\mu)S(\psi_{BA}(t,\mu),
   \psi_{BA}(t,-\lambda)_x)\\
   &=& \int d\mu
   \mu^{-1}\varphi(\mu)\psi_{BA}(t,\mu)\frac{1}{2}\Big(\psi_{BA}(t,-\lambda)-\psi_{BA}(t-2[\mu^{-1}],-\lambda)\Big)
   using (\ref{sepbiaozhun})\\
   &=&\frac{1}{2}\psi_{BA}(t,-\lambda)\Phi(t)-\frac{1}{2}\underline{\int d\mu
   \mu^{-1}\varphi(\mu)\psi_{BA}(t-2[\mu^{-1}],-\lambda)\psi_{BA}(t,\mu)}.
\end{eqnarray*}
Thus we only need to compute the underlied part above. To this end,
with the help of
 $e^{\xi(-2[\mu^{-1}],-\lambda)}=\frac{\mu+\lambda}{\mu-\lambda}$ in (\ref{translationidentity}),we first calculate
\begin{eqnarray}
  &&\psi_{BA}(t-2[\mu^{-1}],-\lambda))\psi_{BA}(t,\mu)\nonumber \\
  &=& (\lambda+\mu)\frac{1}{\mu}\frac{1}{1-\frac{\lambda}{\mu}}
  e^{\xi(t,\mu)-\xi(t,\lambda)}\frac{\tau(t+2[\lambda^{-1}]-2[\mu^{-1}])}{\tau(t)} \nonumber\\
  &=&(\lambda+\mu)[\delta(\lambda,\mu)-\frac{1}{\lambda}\frac{1}{1-\frac{\mu}{\lambda}}]
  e^{\xi(t,\mu)-\xi(t,\lambda)}\frac{\tau(t+2[\lambda^{-1}]-2[\mu^{-1}])}{\tau(t)}\nonumber\\
  &=&-  \psi_{BA}(t+2[\lambda^{-1}],\mu) \psi_{BA}(t,-\lambda)+(\lambda  + \mu) \delta (\lambda ,\mu
  ).\label{zhongyaoguanxi}
\end{eqnarray}
Here, the delta-function is defined as
\begin{equation}\label{delta}
    \delta (\lambda ,\mu )=\frac{1}{\mu}\sum_{n=-\infty}^\infty(\frac{\mu}{\lambda})^n=
    \frac{1}{\lambda}\frac{1}{1-\frac{\mu}{\lambda}}+\frac{1}{\mu}\frac{1}{1-\frac{\lambda}{\mu}}
\end{equation}
and the following property of delta-function is used: \ given a
function $f(z)=\sum_{i=-\infty}^\infty a_i z^i$,
$$f(z)\delta (\lambda ,z )=f(\lambda)\delta (\lambda ,z )$$
as is seen from
$z^i\sum_n(z/\lambda)^n=\lambda^i\sum_n(z/\lambda)^{n+i}$. Thus
taking (\ref{zhongyaoguanxi}) back into the underlined part above,
then
\begin{eqnarray*}
  &&\int d\mu
   \mu^{-1}\varphi(\mu)\psi_{BA}(t-2[\mu^{-1}],-\lambda))\psi_{BA}(t,\mu)\\
   &=& -\int d\mu
   \mu^{-1}\varphi(\mu)\psi_{BA}(t+2[\lambda^{-1}],\mu) \psi_{BA}(t,-\lambda)+\int d\mu
   \mu^{-1}\varphi(\mu) (\lambda  + \mu) \delta (\lambda ,\mu  )\\
   &=& -\Phi(t+2[\lambda^{-1}])\psi_{BA}(t,-\lambda)+the~ term~
   independent~ of ~t.
\end{eqnarray*}
So we get
\begin{equation}\label{bsep5}
    S(\Phi(t),\psi_{BA}(t,-\lambda)_{x}))=\frac{1}{2}\psi_{BA}(t,-\lambda)\big(\Phi(t+2[\lambda^{-1}])+\Phi(t)\big),
\end{equation}
since the definition of SEP up to the term independent of t.

Similarly, we can get the expressions of
$S(\psi_{BA}(t,\lambda),\Phi_x(t))$ and $S(\Phi_1(t),\Phi_{2x}(t))$
by considering $\Phi_x(t)=\int d\lambda
\lambda^{-1}\varphi(\lambda)\psi_{BA}(t,\lambda)_x$, see appendix
\ref{appcorollary3} and \ref{appcorollary4}. Thus we have the
following corollary.
\begin{corollary}\label{corbasicsep}If $\Phi(t),\Phi_1(t),\Phi_2(t)$ are  eigenfunctions of the {\rm BKP} hierarchy, then
\begin{eqnarray}
  &&S(\psi_{BA}(t,\mu),\psi_{BA}(t,-\lambda)_x)=\frac{1}{2}\big(\psi_{BA}(t,-\lambda)
  -\psi_{BA}(t-2[\mu^{-1}],-\lambda) \big)\psi_{BA}(t,\mu)\label{corollary1}, \\
  &&S(\Phi(t),\psi_{BA}(t,-\lambda)_{x})=\frac{1}{2}\psi_{BA}(t,-\lambda)\big(\Phi(t+2[\lambda^{-1}])+\Phi(t)\big)
  \label{corollary2},\\
  &&S(\psi_{BA}(t,\lambda),\Phi_x(t)) =  \frac{1}{2}\psi_{BA}(t,\lambda)\big(\Phi(t)-\Phi(t-2[\lambda^{-1}])\big)\label{corollary3}, \\
  &&S(\Phi_1(t),\Phi_{2x}(t)) = \int\int d\lambda d\mu \lambda^{-1}\mu^{-1}
  \varphi_1(\mu) \varphi_2(\lambda)S\big(\psi_{BA}(t,\mu),\psi_{BA}(t,\lambda)_x\big)\label{corollary4}.
\end{eqnarray}
\end{corollary}
\noindent\textbf{Remark 4}: Note that (\ref{corollary3}) is also
derived by Loirs\cite{LorisWillox99} by a
different method.\\
\noindent\textbf{Remark 5}:According to proposition
\ref{propspectralrep} and corollary \ref{corbasicsep}, we have
\begin{equation}\label{srbkp3}
    \Phi(t)=\int d\lambda
    \lambda^{-1}\psi_{BA}(t,\lambda)\psi_{BA}(t',-\lambda)[\frac{1}{2}\Phi(t'+2[\lambda^{-1}])+\frac{1}{2}\Phi(t')].
\end{equation}

In fact,the inverse of proposition \ref{propspectralrep}  is also
correct and it provides another formulation of the BKP hierarchy,
that is,
\begin{proposition}
Given a function $\psi(t,\lambda)$ which has the form
$\psi(t,\lambda)=e^{\xi(t,\lambda)}\Sigma_{j=0}^\infty\omega_j(t)\lambda^{-j}$
with $\omega_0=1$ and $\xi(t,\lambda)$ as in (\ref{wavefun}),where
multi-time $t=(t_1=x,t_3,\cdots)$ and $\lambda$ is the spectral
parameter,let us assume that $\psi(t,\lambda)$ has the following
spectral representation:
\begin{equation}\label{srbkp5}
    \psi(t,\mu)=\int d\lambda
\lambda^{-1}\psi(t,\lambda)S(t';\lambda,\mu),
\end{equation}
for two arbitrary multi-times $t$ and $t'$,where the function
$S(t;\lambda,\mu)$ is defined such that $\frac{\pa}{\pa
t_1}S(t;\lambda,\mu)=\psi(t,\mu)\psi(t,-\lambda)_{x}$.Then,(\ref{srbkp5})is
equivalent to the Hirota bilinear identity (\ref{bileq}),so in this
way $\psi(t,\lambda)$ becomes BA functions of the associated {\rm
BKP} hierarchy.
\end{proposition}
\textbf{Proof:} The proof for one side of the equivalence that
Hirota bilinear identity (\ref{bileq}) imply the spectral
representation (\ref{srbkp5}), is contained in the proof of
(\ref{srbkp6}). So we only need to show that (\ref{srbkp5}) implies
(\ref{bileq}).To the end, by differentiating both side of
(\ref{srbkp5})w.r.t.$t'_1$, then,
$$0=\pa\psi(t,\lambda)/\pa t'_1=\psi(t',\mu)\int d\lambda
\lambda^{-1}\psi(t,\lambda)\psi(t',-\lambda)_{x'}.$$
 So
$$\int d\lambda
\lambda^{-1}\psi(t,\lambda)\psi(t',-\lambda)\equiv C.$$ By letting
$t'=t$, and considering $\psi(t,\lambda)\psi(t,-\lambda)=1+\mathcal
{O}(\lambda^{-1})$, we have $C=1$. Thus $\psi(t,\lambda)$ satisfies
$\int d\lambda \lambda^{-1}\psi(t,\lambda)\psi(t',-\lambda)=1$,i.e.,
the Hiorta biliner equations of the BKP hierarchy. $\Box$

By now,we have established the SEP method for the BKP
hierarchy,which provides another formulation of the BKP hierarchy.

\section{BSEP}
Based on the useful properties of the SEP given the last section, we
are now in a position to discuss a new potential $\Omega$--BSEP
\cite{LorisWillox99}, which will be used to generate the ``ghost"
flow of the BKP hierarchy in the next section. We first provide
three expressions of $\Omega$ for different eigenfunctions, and then
give their identities.

BSEP is also defined as a function of a pair of BKP eigenfunctions
$\Phi_1$ and $\Phi_2$:
\begin{equation}\label{bkpsep}
    \Omega(\Phi_1,\Phi_2)=S(\Phi_2,\Phi_{1x})-S(\Phi_1,\Phi_{2x}).
\end{equation}
The definition of BSEP can be up to a constant of integration. It is
obvious that $\Omega(\Phi_1,\Phi_2) =-\Omega(\Phi_2,\Phi_1)$ and
that $\Omega(\Phi,1)=\Phi$ (since 1 is an eigenfunction). So
according to (\ref{sepbiaozhun})and (\ref{zhongyaoguanxi}),we have
\begin{eqnarray}
  \Omega\big(\psi_{BA}(t,\mu),\psi_{BA}(t,-\lambda)\big)&=& -\psi_{BA}(t+2[\lambda^{-1}],\mu) \psi_{BA}(t,-\lambda)+\frac{1}{2}(\lambda  +
\mu) \delta (\lambda ,\mu )\nonumber \\
  &=& \psi_{BA}(t,\mu) \psi_{BA}(t-2[\mu^{-1}],-\lambda)-\frac{1}{2}(\lambda  +
\mu) \delta (\lambda ,\mu ) \label{bkpsepp1}.
\end{eqnarray}
\textbf{Remark 6}: As the definition of BSEP can be up to the term
independent of $t$, we can omit the terms independent of $t$ in
(\ref{bkpsepp1}). That is,
\begin{equation}\label{bkpsep1}
    \Omega\big(\psi_{BA}(t,\mu),\psi_{BA}(t,-\lambda)\big)= -\psi_{BA}(t+2[\lambda^{-1}],\mu) \psi_{BA}(t,-\lambda).
\end{equation}

We would like to mention there is another expression for
$\Omega(\psi_{BA}(t,\mu),\psi_{BA}(t,-\lambda))$,i.e.,
\begin{equation}\label{anotherexpression}
    \Omega\big(\psi_{BA}(t,\mu),\psi_{BA}(t,-\lambda)\big)=-\frac{\mathcal {X}(\lambda ,\mu )\tau(t)}{\tau(t)}.
\end{equation}
Here the vertex operator \cite{DJKM} is defined as follows,
\begin{eqnarray}
  \mathcal {X}(\lambda ,\mu )&\equiv& :e^{\theta {\rm{(}}\lambda {\rm{)}}} ::e^{ - \theta {\rm{(}}\mu
{\rm{)}}} :
 = e^{\xi {\rm{(t + 2[}}\lambda ^{ - 1} {\rm{],}}\mu {\rm{) -
}}\xi {\rm{(t,}}\lambda {\rm{)}}} e^{\sum\limits_1^\infty
{\frac{2}{{2l - 1}}(\lambda ^{ - (2l - 1)}  - \mu ^{ - (2l - 1)}
)\frac{\partial }{{\partial t_{2l - 1} }}} }\nonumber\\
   &=&- e^{\xi {\rm{(t,}}\mu {\rm{) - }}\xi {\rm{(t - 2[}}\mu ^{ - 1} {\rm{],}}\lambda {\rm{)}}} e^{\sum\limits_1^\infty  {\frac{2}{{2l - 1}}(\lambda ^{ - (2l - 1)}  - \mu ^{ - (2l - 1)} )\frac{\partial }{{\partial t_{2l - 1} }}} }
   +(\lambda  + \mu) \delta (\lambda ,\mu ) \label{vop}
\end{eqnarray}
where
\begin{equation}\label{sita}
   \theta(\lambda)\equiv
   -\sum\limits_{l=1}^\infty\lambda^{2l-1}t_{2l-1}+\sum\limits_{l=1}^\infty\frac{1}{2l-1}\lambda^{-(2l-1)}\frac{\partial
   }{\partial t_{2l-1}}
\end{equation}
the columns $:\cdots:$ indicate Wick Normal ordering w.r.t the
creation/annihilation "modes"$t_l$ and $\frac{\partial }{\partial
t_l}$,respectively.  Thus according to the definition of the Vertex
operator (\ref{vop} ) and the wave function (\ref{wavefun}) and
(\ref{tau}),we can easily get
\begin{eqnarray}
  \frac{\mathcal {X}(\lambda ,\mu )\tau(t)}{\tau(t)} &=& \psi_{BA}(t+2[\lambda^{-1}],\mu) \psi_{BA}(t,-\lambda)
   \nonumber \\
 &=&-  \psi_{BA}(t,\mu) \psi_{BA}(t-2[\mu^{-1}],-\lambda)+(\lambda  + \mu) \delta (\lambda ,\mu ).\label{ba2}
\end{eqnarray}
So (\ref{anotherexpression}) is true.

As for $\Omega(\Phi(t),\psi_{BA}(t,-\lambda))$, according to the
definition of $\Omega$, two identities (\ref{corollary2}) and
(\ref{corollary3}), then
 $\Omega(\Phi(t),\psi_{BA}(t,-\lambda))$ can expressed  by the form of
\begin{eqnarray}
  \Omega(\Phi(t),\psi_{BA}(t,-\lambda)) &=& S(\psi_{BA}(t,-\lambda),\Phi_x(t))
    -S(\Phi(t),\psi_{BA}(t,-\lambda)_{x})) \nonumber \\
  &=&
  \frac{1}{2}(\Phi(t)-\Phi(t+2[\lambda^{-1}]))\psi_{BA}(t,-\lambda)-\frac{1}{2}\psi_{BA}(t,-\lambda)[\Phi(t+2[\lambda^{-1}])+\Phi(t)]
\nonumber
  \\
  &=&-\psi_{BA}(t,-\lambda)\Phi(t+2[\lambda^{-1}]). \label{Omegaphipsinegativelambda}
\end{eqnarray}
Note that (\ref{Omegaphipsinegativelambda}) implies (\ref{bkpsep1}) in Remark 6 as we expected.\\
\textbf{Remark 7}:
 In fact,with the help of spectral representation (\ref{srbkp1}) for
the BKP hierarchy  and the expression for
$\Omega(\psi_{BA}(t,\mu),\psi_{BA}(t,-\lambda))$ (\ref{bkpsep1}) ,
$\Omega(\Phi(t),\psi_{BA}(t,-\lambda))$ is derived alternatively  as
\begin{eqnarray*}
   &&\Omega(\Phi(t),\psi_{BA}(t,-\lambda)) \\
 &=& \int d\mu\mu^{-1}\varphi(\mu)\Omega(\psi_{BA}(t,\mu),\psi_{BA}(t,-\lambda)) \\
  &=&-\int d\mu\mu^{-1}\varphi(\mu)\psi_{BA}(t+2[\lambda^{-1}],\mu)
  \psi_{BA}(t,-\lambda)   \\
  &=&-\Phi(t+2[\lambda^{-1}])\psi_{BA}(t,-\lambda).
\end{eqnarray*}
We further show a more general $\Omega$ of eigenfunctions $\Phi_1$
and $\Phi_2$,
\begin{equation}\label{bkpsep2}
    \Omega(\Phi_1,\Phi_2)=\int\int d\lambda d\mu \lambda^{-1}\mu^{-1}\varphi_1(\mu) \varphi_2(\lambda)\Omega(\psi_{BA}(t,\mu),\psi_{BA}(t,\lambda)).
\end{equation}

Next,we would like to show three identities on $\Omega$ of the BKP
hierarchy.
\begin{lemma} For the {\rm BKP} hierarchy,
\begin{equation}\label{shdo}
    \Delta_z(\psi_{BA}(t,\mu) \psi_{BA}(t-2[\mu^{-1}],\lambda))=\psi_{BA}(t,\lambda)
    \psi_{BA}(t-2[z^{-1}],\mu)-\psi_{BA}(t,\mu)
    \psi_{BA}(t-2[z^{-1}],\lambda)
\end{equation}
where
$\Delta_z=e^{-\sum\limits_{l=1}^\infty\frac{2}{2l-1}z^{-(2l-1)}\frac{\partial
   }{\partial t_{2l-1}}}-1$ is a shift-difference operator.
\end{lemma}
{\bf Proof:} First of all, we move all terms in right side hand  of
(\ref{shdo}) to the left,
 take $\psi_{BA}(t,\lambda)$ in (\ref{wavefun}) and (\ref{tau}) into it, then
\begin{eqnarray*}
  (\ref{shdo}) {\rm holds} &\Leftrightarrow&\psi_{BA}(t-2[z^{-1}],\mu) \psi_{BA}(t-2[\mu^{-1}]-2[z^{-1}],\lambda)-\psi_{BA}(t,\mu)
  \psi_{BA}(t-2[\mu^{-1}],\lambda)\\
  &&+\psi_{BA}(t,\mu)
    \psi_{BA}(t-2[z^{-1}],\lambda)-\psi_{BA}(t,\lambda)
    \psi_{BA}(t-2[z^{-1}],\mu)=0\\
   &\Leftrightarrow&\frac{(z-\mu)(z-\lambda)(\mu-\lambda)}{(z+\mu)(z+\lambda)(\mu+\lambda)}
   \frac{\tau(t-2[\mu^{-1}]-2[z^{-1}]-2[\lambda^{-1}])}{\tau(t-2[z^{-1}])}\\
   &&-\frac{\mu-\lambda}{\mu+\lambda}
   \frac{\tau(t-2[\mu^{-1}]-2[\lambda^{-1}])}{\tau(t)}
   +\frac{z-\lambda}{z+\lambda}
   \frac{\tau(t-2[\mu^{-1}])}{\tau(t)}
   \frac{\tau(t-2[z^{-1}]-2[\lambda^{-1}])}{\tau(t-2[z^{-1}])}\\
   &&-\frac{z-\mu}{z+\mu}
   \frac{\tau(t-2[\lambda^{-1}])}{\tau(t)}
   \frac{\tau(t-2[z^{-1}]-2[\mu^{-1}])}{\tau(t-2[z^{-1}])}=0
\ using (\ref{translationidentity}),\ removing\
e^{\xi(t,\lambda)+\xi(t,\mu)}
   \\
   &\Leftrightarrow&\frac{(z-\mu)(z-\lambda)(\mu-\lambda)}{(z+\mu)(z+\lambda)(\mu+\lambda)}
   \tau(t-2[\mu^{-1}]-2[z^{-1}]-2[\lambda^{-1}])\tau(t)\\
   &&-\frac{\mu-\lambda}{\mu+\lambda}\tau(t-2[\mu^{-1}]-2[\lambda^{-1}])\tau(t-2[z^{-1}])
   +\frac{z-\lambda}{z+\lambda}
   \tau(t-2[\mu^{-1}])\tau(t-2[z^{-1}]-2[\lambda^{-1}])\\
   &&-\frac{z-\mu}{z+\mu}\tau(t-2[\lambda^{-1}])\tau(t-2[z^{-1}]-2[\mu^{-1}])=0
\  multiplying \ \tau(t)\tau(t-2[z^{-1}])
\\
   &\Leftrightarrow&\frac{(\mu-z)(\mu-\lambda)}{(\mu+z)(\mu+\lambda)}
\tau(t)\tau(t-2[\mu^{-1}]-2[z^{-1}]-2[\lambda^{-1}])\\
&&+\frac{(z+\lambda)(z-\mu)}{(z-\lambda)(z+\mu)}
\tau(t-2[z^{-1}]-2[\mu^{-1}])\tau(t-2[\lambda^{-1}])\\
&&+\frac{(\lambda-\mu)(\lambda+z)}{(\lambda+\mu)(\lambda-z)}
\tau(t-2[\mu^{-1}]-2[\lambda^{-1}])\tau(t-2[z^{-1}])\\
&&- \tau(t-2[\mu^{-1}])\tau(t-2[z^{-1}]-2[\lambda^{-1}])=0.\
multiplying \ \ \frac{\lambda+z}{\lambda-z}
\end{eqnarray*}
For convenience, denote the left hand side of above equality  by
$C$.  Secondly, we  shall prove indeed $C=0$ from the Fay
identity(\ref{Fayeq}) of the  BKP hierarchy, thus (\ref{shdo}) is
proved. To this end, by letting $s_0=0$ in Fay
identity(\ref{Fayeq}), then
\begin{eqnarray*}
&&\frac{(s_1+s_2)(s_1+s_3)}{(s_1-s_2)(s_1-s_3)}
\tau(t+2[s_2]+2[s_3])\tau(t+2[s_1])\\
&&+\frac{(s_2+s_3)(s_2+s_1)}{(s_2-s_3)(s_2-s_1)}
\tau(t+2[s_3]+2[s_1])\tau(t+2[s_2])\\
&&+\frac{(s_3+s_1)(s_3+s_2)}{(s_3-s_1)(s_3-s_2)}
\tau(t+2[s_1]+2[s_2])\tau(t+2[s_3])\\
&&-\tau(t+2[s_1]+2[s_2]+2[s_3])\tau(t)=0.
\end{eqnarray*}
Then,after shifting $t\mapsto t-2[s_2]-2[s_3]$ and letting
$[s_1]\mapsto-[s_1]$ in above equation, it becomes
\begin{eqnarray*}
&&\frac{(s_1-s_2)(s_1-s_3)}{(s_1+s_2)(s_1+s_3)}
\tau(t)\tau(t-2[s_1]-2[s_2]-2[s_3])\\
&&+\frac{(s_2+s_3)(s_2-s_1)}{(s_2-s_3)(s_2+s_1)}
\tau(t-2[s_2]-2[s_1])\tau(t-2[s_3])\\
&&+\frac{(s_3-s_1)(s_3+s_2)}{(s_3+s_1)(s_3-s_2)}
\tau(t-2[s_1]-2[s_3])\tau(t-2[s_2])\\
&&-\tau(t-2[s_1])\tau(t-2[s_2]-2[s_3])=0.
\end{eqnarray*}
At last, setting $s_1=\mu^{-1},s_2=z^{-1},s_3=\lambda^{-1}$,we have
\begin{eqnarray*}
&&\frac{(\mu-z)(\mu-\lambda)}{(\mu+z)(\mu+\lambda)}
\tau(t)\tau(t-2[\mu^{-1}]-2[z^{-1}]-2[\lambda^{-1}])\\
&&+\frac{(z+\lambda)(z-\mu)}{(z-\lambda)(z+\mu)}
\tau(t-2[z^{-1}]-2[\mu^{-1}])\tau(t-2[\lambda^{-1}])\\
&&+\frac{(\lambda-\mu)(\lambda+z)}{(\lambda+\mu)(\lambda-z)}
\tau(t-2[\mu^{-1}]-2[\lambda^{-1}])\tau(t-2[z^{-1}])\\
&&- \tau(t-2[\mu^{-1}])\tau(t-2[z^{-1}]-2[\lambda^{-1}])=0,
\end{eqnarray*}
i.e., $C=0$, as we claimed before. This is the end of the
proof.\hspace{4cm} $\Box$

After the preparation above,now we can give two important identities
of the BSEP below.
\begin{proposition}
Under shift of the times $t$ of the {\rm BKP} hierarchy, {\rm BSEP}
obeys
\begin{eqnarray}
        &&\Omega(\Phi_1(t-2[z^{-1}]),\Phi_2(t-2[z^{-1}]))-\Omega(\Phi_1(t),\Phi_2(t))\nonumber\\
        &&=\Phi_1(t-2[z^{-1}])\Phi_2(t)-\Phi_1(t)\Phi_2(t-2[z^{-1}]),\label{bkpsep3}\\
        &&\Omega(\Phi_1(t+2[z^{-1}]),\Phi_2(t+2[z^{-1}]))-\Omega(\Phi_1(t),\Phi_2(t))\nonumber\\
        &&=\Phi_1(t+2[z^{-1}])\Phi_2(t)-\Phi_1(t)\Phi_2(t+2[z^{-1}]).\label{bkpsep4}
      \end{eqnarray}
\end{proposition}
\textbf{Proof:} By a straightforward calculation, then
\begin{eqnarray*}
\Delta_z\Omega(\Phi_1,\Phi_2)&=&\int\int d\lambda d\mu
\lambda^{-1}\mu^{-1}\varphi_1(\mu)
\varphi_2(\lambda)\Delta_z\Omega\big(\psi_{BA}(t,\mu),\psi_{BA}(t,\lambda) \big)\ \ using(\ref{bkpsep2})   \\
  &=&\int\int d\lambda d\mu
\lambda^{-1}\mu^{-1}\varphi_1(\mu)
\varphi_2(\lambda)\big(\psi_{BA}(t,\lambda)
    \psi_{BA}(t-2[z^{-1}],\mu)\\
    &&-\psi_{BA}(t,\mu)
    \psi_{BA}(t-2[z^{-1}],\lambda)\big), \ \ using\ (\ref{bkpsep1})\ and\ (\ref{shdo})  \\
    &=&\Phi_1(t-2[z^{-1}])\Phi_2(t)-\Phi_1(t)\Phi_2(t-2[z^{-1}]).~using~(\ref{srbkp1})
\end{eqnarray*}
So (\ref{bkpsep3}) is proved. By shift $t\mapsto t+ 2[z^{-1}]$,
(\ref{bkpsep4}) is derived from (\ref{bkpsep3}).
$\Box$\\
\textbf{Remark 8}: In fact,these identities above have been given in
Loris' paper \cite{LorisWillox99}, but here we give another proof
and our proof is much easier.
\section{``Ghost" Symmetry}
After the preparation above, now we can define the ``ghost" symmetry
flows generated by the BSEP through its action on the Lax operator.
We shall further show its actions on the dressing operator,
eigenfunction $\Phi(t)$ and $\tau$ function.

Given a set of eigenfunctions
$\Phi_{1a},\Phi_{2a},a\in\{\alpha\}$,the ``ghost" symmetry of the
BKP hierarchy is defined in the following way
\begin{equation}\label{ghsym1}
    \pa_\alpha
    L\equiv[\sum_{a\in\{\alpha\}}(\Phi_{2a}\pa^{-1}\Phi_{1a,x}-\Phi_{1a}\pa^{-1}\Phi_{2a,x}),L];
\pa_\alpha W
\equiv\sum_{a\in\{\alpha\}}(\Phi_{2a}\pa^{-1}\Phi_{1a,x}-\Phi_{1a}\pa^{-1}\Phi_{2a,x})W
\end{equation}
Next,we need to show that,the definition above is consistent with
the BKP constraint (\ref{constr}) and $\pa_\alpha$ commutes with
$\pa_{t_{2n+1}}$. In other words, $\pa_\alpha$ is indeed a kind of
symmetry flow of the BKP hierarchy. For simplicity  in the next
context, we introduce an operator $ A=
\sum_{a\in\{\alpha\}}(\Phi_{2a}\pa^{-1}\Phi_{1a,x}-\Phi_{1a}\pa^{-1}\Phi_{2a,x})$.
\begin{proposition}
$\pa_\alpha$ is consistent with the {\rm BKP} constraint
(\ref{constr}),i.e.$(\pa_\alpha L^*)\pa+\pa (\pa_\alpha L)=0$.
\end{proposition}
\noindent\textbf{Proof:} According to the definition of $A$, and
using a identity
 $\partial^{-1} f=f \partial^{-1}- \partial^{-1} f_x \partial^{-1}$, we have
\begin{eqnarray*}
  A^*\pa+\pa A &=&
  \sum_{a\in\{\alpha\}}(\Phi_{2a,x}\pa^{-1}\Phi_{1a}\pa-\Phi_{1a,x}\pa^{-1}\Phi_{2a}\pa
  +\pa\Phi_{2a}\pa^{-1}\Phi_{1a,x}-\pa\Phi_{1a}\pa^{-1}\Phi_{2a,x}) \\
  &=&
  \sum_{a\in\{\alpha\}}(\Phi_{2a,x}\Phi_{1a}-\Phi_{2a,x}\pa^{-1}\Phi_{1a,x}
  -\Phi_{1a,x}\Phi_{2a}+\Phi_{1a,x}\pa^{-1}\Phi_{2a,x}\\
  &&+\Phi_{2a}\Phi_{1a,x}+\Phi_{2a,x}\pa^{-1}\Phi_{1a,x}
  -\Phi_{1a}\Phi_{2a,x}-\Phi_{1a,x}\pa^{-1}\Phi_{2a,x})\\
  &=&0.
\end{eqnarray*}
Furthermore, using the definition of $\partial_{\alpha}$, a simple
computation leads to
\begin{eqnarray*}
  (\pa_\alpha L^*)\pa+\pa (\pa_\alpha L) &=& [A,L]^*\pa+\pa[A,L]= -[A^*,L^*]\pa+\pa[A,L]\\
  &=& -\pa L \pa^{-1}A^*\pa+A^*\pa L \pa^{-1}\pa+\pa[A,L]\\
  &=& -\pa (L \pa^{-1}A^*\pa-\pa^{-1}A^*\pa L)+\pa[A,L]\\
  &=& \pa[\pa^{-1}A^*\pa,L] +\pa[A,L]= \pa[\pa^{-1}A^*\pa+A,L]=0,
\end{eqnarray*}
because of the above identity on $A$. This means $\partial_{\alpha}
L^*$ is consistent with BKP constraint (\ref{constr}). $\Box$
\begin{proposition}
$\pa_\alpha$ commutes with $\pa_{t_{2n+1}}$.
\end{proposition}
\noindent\textbf{Proof}: We first claim the following equations
\begin{equation}\label{zhengming3}
    \pa_\alpha B_{2n+1}-\pa_{t_{2n+1}} A=[A,B_{2n+1}]
\end{equation}
hold for $A$ and $\pa_\alpha $, which will be proved latter. With
the help of above equation, a simple calculation infers
\begin{eqnarray*}
  [\pa_{t_{2n+1}},\pa_\alpha]L &=& \pa_{t_{2n+1}}([A,L])-\pa_\alpha([B_{2n+1},L]) \\
   &=& [\pa_{t_{2n+1}}A,L]+[A,[B_{2n+1},L]]-[\pa_\alpha
   B_{2n+1},L]-[B_{2n+1},[A,L]]\\
   &=& [\pa_{t_{2n+1}}A-\pa_\alpha B_{2n+1}+[A,B_{2n+1}],L] \ using \ \ Jacobi \ \ identity \\
   &=&0,
\end{eqnarray*}
which shows $\pa_\alpha$ commutes with $\pa_{t_{2n+1}}$. Therefore,
the remaining part of the proof is to show our claimed statement
(\ref{zhengming3}). First of all, the definition of the ``ghost"
flows $\pa_\alpha L=[A,L]$ implies obviously $\pa_\alpha
L^{2n+1}=[A,L^{2n+1}]$ .Thus,we have
\begin{equation}\label{zhengming1}
    \pa_\alpha B_{2n+1}=([A,L^{2n+1}])_+=([A,B_{2n+1}])_+
\end{equation}
Secondly, the derivative of $A$ with respect to $t_{2n+1}$ is given
by
\begin{eqnarray*}
\pa_{t_{2n+1}}A =\sum_{a\in\{\alpha\}}\big(
(\pa_{t_{2n+1}}\Phi_{2a})\pa^{-1}\Phi_{1a,x}-(\pa_{t_{2n+1}}\Phi_{1a})\pa^{-1}\Phi_{2a,x}\big)\\
+\sum_{a\in\{\alpha\}}\big( \Phi_{2a}\pa^{-1}(\pa_{t_{2n+1}}
\Phi_{1a,x})-\Phi_{1a}\pa^{-1}(\pa_{t_{2n+1}}\Phi_{2a,x})\big).
\end{eqnarray*}
Taking (\ref{eigen}) into it, then
\begin{eqnarray*}
\pa_{t_{2n+1}}A =\sum_{a\in\{\alpha\}}\big(
B_{2n+1}(\Phi_{2a})\pa^{-1}\Phi_{1a,x}-B_{2n+1}(\Phi_{1a})\pa^{-1}\Phi_{2a,x}\big)\\
-\sum_{a\in\{\alpha\}}\big( \Phi_{2a}\pa^{-1}B_{2n+1}^*(
\Phi_{1a,x})-\Phi_{1a}\pa^{-1}B_{2n+1}^*(\Phi_{2a,x})\big).
\end{eqnarray*}
Note $\Phi_{1a,x} $ and  $\Phi_{2a,x} $ are two adjoint
eigenfunctions. Furthermore\footnote{Here the relation
$(F_+\pa^{-1})_-=F_{[0]} \pa^{-1} $($F$ is a pseudo-differential
operator) is used.}\    \footnote{with the relation
$(\pa^{-1}F_+)_-=\pa^{-1}(F^*)_{[0]}$($F$ is a pseudo-differential
operator)},
\begin{eqnarray*}
\pa_{t_{2n+1}}A &=&\Big(B_{2n+1} \sum_{a\in\{\alpha\}}\big(
\Phi_{2a}\pa^{-1}\Phi_{1a,x}-\Phi_{1a}\pa^{-1}\Phi_{2a,x}\big)\Big)_-\\
&-&\Big(\sum_{a\in\{\alpha\}}\big(
\Phi_{2a}\pa^{-1} \Phi_{1a,x}-\Phi_{1a}\pa^{-1}\Phi_{2a,x})\big)B_{2n+1} \Big)_-\\
&=&\big(B_{2n+1}A \big)_- - \big(A B_{2n+1}\big)_-.
\end{eqnarray*}
Thus we have
\begin{equation}\label{zhengming2}
    \pa_{t_{2n+1}} A=-([A,B_{2n+1}])_-,
\end{equation}
At last,  according to  (\ref{zhengming1})$- $(\ref{zhengming2}) ,
(\ref{zhengming3}) is obtained. \hspace{2cm}$\Box$

Next,let's see the action of the above ``ghost" flows on the
eigenfunctions $\Phi$:
\begin{proposition}The ``ghost" symmetry is the compatible condition of
the linear problems
\begin{eqnarray}
  \pa_{t_{2n+1}}\Phi &=&B_{2n+1}(\Phi), \\
  \pa_\alpha \Phi &=&\frac{1}{2}
\sum\limits_{a\in\{\alpha\}}(\Phi_{2a}\Omega(\Phi_{1a},\Phi)-\Phi_{1a}\Omega(\Phi_{2a},\Phi)).\label{ghsym2}
\end{eqnarray}
\end{proposition}
\noindent\textbf{Proof:} The main idea of the proof is to use
(\ref{zhengming3}), which is equivalent to ghost symmetry flow
(\ref{ghsym1}), to get the commutativity of the flows
$\pa_\alpha\pa_{t_{2n+1}} \Phi =\pa_{t_{2n+1}}\pa_\alpha\Phi$. So,
according to (\ref{lemma1}),we can rewrite (\ref{ghsym2}) into
\begin{eqnarray}
\pa_\alpha \Phi &=& \frac{1}{2}
\sum\limits_{a\in\{\alpha\}}\Phi_{2a}(S(\Phi,\Phi_{1ax})-S(\Phi_{1a},\Phi_x))-(1\leftrightarrow2)\nonumber \\
 &=&
\sum\limits_{a\in\{\alpha\}}(\Phi_{2a}S(\Phi,\Phi_{1ax})-\frac{1}{2}\Phi_{1a}\Phi_{2a}\Phi)-(1\leftrightarrow2)\nonumber\\
 &=&
\sum\limits_{a\in\{\alpha\}}(\Phi_{2a}S(\Phi,\Phi_{1ax})-\Phi_{1a}S(\Phi,\Phi_{2ax})),\label{proof}
\end{eqnarray}
and then
\begin{eqnarray}
&&\pa_{t_{2n+1}}(\pa_\alpha \Phi) \nonumber \\
 &=&
\sum\limits_{a\in\{\alpha\}}\{(\pa_{t_{2n+1}}\Phi_{2a})S(\Phi,\Phi_{1ax})
-(\pa_{t_{2n+1}}\Phi_{1a})S(\Phi,\Phi_{2ax}) \nonumber\\
&&+\Phi_{2a}Res(\pa^{-1}\Phi_{1ax}B_{2n+1}\Phi\pa^{-1})
-\Phi_{1a}Res(\pa^{-1}\Phi_{2ax}B_{2n+1}\Phi\pa^{-1})\}\nonumber\\
 &=&
\sum\limits_{a\in\{\alpha\}}\{(\pa_{t_{2n+1}}\Phi_{2a})S(\Phi,\Phi_{1ax})
-(\pa_{t_{2n+1}}\Phi_{1a})S(\Phi,\Phi_{2ax})\}+Res(AB_{2n+1}\Phi\pa^{-1}).\label{xinjia}
\end{eqnarray}
By a tedious but straightforward calculation, we have(see appendix
\ref{appflowcommutativity2})
\begin{eqnarray}
Res(AB_{2n+1}\Phi\pa^{-1})=\pa_\alpha(\pa_{t_{2n+1}}\Phi)-\sum\limits_{a\in\{\alpha\}}((\pa_{t_{2n+1}}\Phi_{2a})S(\Phi,\Phi_{1a,x})
-(\pa_{t_{2n+1}}\Phi_{1a})S(\Phi,\Phi_{2a,x})).
 \label{flowcommutativity2}
\end{eqnarray}
Thus by substituting (\ref{flowcommutativity2}) into
(\ref{xinjia}),we get
\begin{equation*}
\pa_\alpha(\pa_{t_{2n+1}}\Phi)=\pa_{t_{2n+1}}(\pa_\alpha\Phi).
\hspace{7cm}\Box
\end{equation*}

 We now consider the commutativity of two ``ghost" symmetries generated by different pairs of eigenfunctions
$\{\Phi_{1a},\Phi_{2a}\}_{a\in\{\alpha\}}$ and
$\{\Phi_{1b},\Phi_{2b}\}_{b\in\{\beta\}}$, and their corresponding
flows are $\pa_{\alpha}L=[A,L]$ and  $\pa_{\beta}L=[A',L]$. Here
$A=\sum_{a\in\{\alpha\}}(\Phi_{2a}\pa^{-1}\Phi_{1a,x}-\Phi_{1a}\pa^{-1}\Phi_{2a,x})
$ as
before,$A'=\sum_{b\in\{\beta\}}\Phi_{2b}(\pa^{-1}\Phi_{1b,x}-\Phi_{1b}\pa^{-1}\Phi_{2b,x})$.

\begin{proposition}
If  two ``ghost" symmetry flows $\pa_{\alpha}$ and $\pa_{\beta}$ are
generated by $A$ and $A'$ above, then $[\pa_\alpha,\pa_\beta]=0$.
\end{proposition}
\noindent{\bf Proof}: By using the relation
\begin{equation}\label{guanxishi}
    f_1\pa^{-1}g_1f_2\pa^{-1}g_2=f_1S(f_2,g_1)\pa^{-1}g_2-f_1\pa^{-1}S(f_2,g_1)g_2,
\end{equation}
then,
\begin{eqnarray*}
  AA' &=& \sum_{a,b}(\Phi_{2a}\pa^{-1}\Phi_{1a,x}-\Phi_{1a}\pa^{-1}\Phi_{2a,x} )(\Phi_{2b}\pa^{-1}\Phi_{1b,x}-\Phi_{1b}\pa^{-1}\Phi_{2b,x})\\
   &=&\sum_{a,b}\Big(\Phi_{2a}\pa^{-1}\Phi_{1a,x}\Phi_{2b}\pa^{-1}\Phi_{1b,x}-\Phi_{2a}\pa^{-1}\Phi_{1a,x}\Phi_{1b}\pa^{-1}\Phi_{2b,x}\\
   &&-\Phi_{1a}\pa^{-1}\Phi_{2a,x}\Phi_{2b}\pa^{-1}\Phi_{1b,x}+\Phi_{1a}\pa^{-1}\Phi_{2a,x}\Phi_{1b}\pa^{-1}\Phi_{2b,x}\Big)\\
   &=&\sum_{a,b}\Big(\Phi_{2a}S(\Phi_{2b},\Phi_{1a,x})\pa^{-1}\Phi_{1b,x}-\Phi_{2a}\pa^{-1}S(\Phi_{2b},\Phi_{1a,x})\Phi_{1b,x}\\
   &&-\Phi_{2a}S(\Phi_{1b},\Phi_{1a,x})\pa^{-1}\Phi_{2b,x}+\Phi_{2a}\pa^{-1}S(\Phi_{1b},\Phi_{1a,x})\Phi_{2b,x}\\
   &&-\Phi_{1a}S(\Phi_{2b},\Phi_{2a,x})\pa^{-1}\Phi_{1b,x}+\Phi_{1a}\pa^{-1}S(\Phi_{2b},\Phi_{2a,x})\Phi_{1b,x}\\
   &&+\Phi_{1a}S(\Phi_{1b},\Phi_{2a,x})\pa^{-1}\Phi_{2b,x}-\Phi_{1a}\pa^{-1}S(\Phi_{1b},\Phi_{2a,x})\Phi_{2b,x}\Big).\\
\end{eqnarray*}
Collecting terms in $AA'$ according to
$\pa^{-1}\Phi_{1b,x}$,$\pa^{-1}\Phi_{2b,x}$, $\Phi_{2a} \pa^{-1}$
and $\Phi_{1a} \pa^{-1}$ in order, then using (\ref{proof}), we have
\begin{eqnarray*}
   AA'&=&\sum_{a,b}\Big(\Phi_{2a}S(\Phi_{2b},\Phi_{1a,x})-\Phi_{1a}S(\Phi_{2b},\Phi_{2a,x}))\pa^{-1}\Phi_{1b,x}\\
   &&-(\Phi_{2a}S(\Phi_{1b},\Phi_{1a,x})-\Phi_{1a}S(\Phi_{1b},\Phi_{2a,x}))\pa^{-1}\Phi_{2b,x}\\
   &&+\Phi_{2a}\pa^{-1}(S(\Phi_{1b},\Phi_{1a,x})\Phi_{2b,x}-S(\Phi_{2b},\Phi_{1a,x})\Phi_{1b,x})\\
   &&+\Phi_{1a}\pa^{-1}(S(\Phi_{2b},\Phi_{2a,x})\Phi_{1b,x}-S(\Phi_{1b},\Phi_{2a,x})\Phi_{2b,x})\Big)\\
   &=&\sum_{b}\Big((\pa_\alpha\Phi_{2b})\pa^{-1}\Phi_{1b,x}-(\pa_\alpha\Phi_{1b})\pa^{-1}\Phi_{2b,x}\Big)\\
   &&+\sum_{a}\Big(-\Phi_{2a}\pa^{-1}(\pa_\beta\Phi_{1a,x})+\Phi_{1a}\pa^{-1}(\pa_\beta\Phi_{2a,x})\Big).
\end{eqnarray*}
So
\begin{eqnarray*}
  &&[A,A']\\
  &=&\sum_{b}\{(\pa_\alpha\Phi_{2b})\pa^{-1}\Phi_{1b,x}-(\pa_\alpha\Phi_{1b})\pa^{-1}\Phi_{2b,x}\}+\sum_{a}\{-\Phi_{2a}\pa^{-1}(\pa_\beta\Phi_{1a,x})+\Phi_{1a}\pa^{-1}(\pa_\beta\Phi_{2a,x})\}\\
  &&+\sum_{a}\{-(\pa_\beta\Phi_{2a})\pa^{-1}\Phi_{1a,x}+(\pa_\beta\Phi_{1a})\pa^{-1}\Phi_{2a,x}\}+\sum_{b}\{\Phi_{2b}\pa^{-1}(\pa_\alpha\Phi_{1b,x})-\Phi_{1b}\pa^{-1}(\pa_\alpha\Phi_{2b,x})\}\\
  &=&\sum_{a}\{-\Phi_{2a}\pa^{-1}(\pa_\beta\Phi_{1a,x})+\Phi_{1a}\pa^{-1}(\pa_\beta\Phi_{2a,x})-(\pa_\beta\Phi_{2a})\pa^{-1}\Phi_{1a,x}+(\pa_\beta\Phi_{1a})\pa^{-1}\Phi_{2a,x}\}\\
  &&+\sum_{b}\{\Phi_{2b}\pa^{-1}(\pa_\alpha\Phi_{1b,x})-\Phi_{1b}\pa^{-1}(\pa_\alpha\Phi_{2b,x})+(\pa_\alpha\Phi_{2b})\pa^{-1}\Phi_{1b,x}-(\pa_\alpha\Phi_{1b})\pa^{-1}\Phi_{2b,x})\}\\
  &=&-\pa_\beta A+\pa_\alpha A'.
\end{eqnarray*}
Hence,
\begin{eqnarray*}
  [\pa_\alpha,\pa_\beta]L&=& \pa_\alpha [A',L]-\pa_\beta[A,L] \\
   &=&[\pa_\alpha A'-\pa_\beta A,L]+[A',[A,L]]-[A,[A',L]]\\
   &=&[\pa_\alpha A'-\pa_\beta A+[A',A],L]=0. \hspace{7cm}\Box
\end{eqnarray*}
At last,let us see the action of ``ghost" flow on the $\tau$
function.
\begin{proposition}
\begin{equation}\label{loris}
    \pa_\alpha\tau(t)=\frac{1}{2}\sum_{a\in\{\alpha\}}\Omega(\Phi_{2a}(t),\Phi_{1a}(t))\tau(t).
\end{equation}
\end{proposition}
\noindent{\bf Proof}: Since $\psi_{BA}(t,\lambda)$ is also an
eigenfunction, so (\ref{ghsym2}) implies
\begin{eqnarray*}
  \pa_\alpha \psi_{BA}(t,\lambda)& =&\frac{1}{2}
\sum_{a\in\{\alpha\}}[\Phi_{2a}(t)\Omega(\Phi_{1a}(t),\psi_{BA}(t,\lambda))-\Phi_{1a}(t)\Omega(\Phi_{2a}(t),\psi_{BA}(t,\lambda)) ]\\
  &=&\frac{1}{2}
\sum_{a\in\{\alpha\}}[-\Phi_{2a}(t)\Phi_{1a}(t-2[\lambda^{-1}])+\Phi_{1a}(t)\Phi_{2a}(t-2[\lambda^{-1}])]\psi_{BA}(t,\lambda)
~using~(\ref{Omegaphipsinegativelambda})\\
&=&\frac{1}{2}
\sum_{a\in\{\alpha\}}\Delta_\lambda\Omega(\Phi_{2a}(t),\Phi_{1a}(t))\psi_{BA}(t,\lambda).
~using~(\ref{bkpsep3})
\end{eqnarray*}
So we have
$\pa_\alpha\tau(t)=\frac{1}{2}\sum_{a\in\{\alpha\}}\Omega(\Phi_{2a}(t),\Phi_{1a}(t))\tau(t)
$
using the expression of $\psi_{BA}(t,\lambda)$ in (\ref{wavefun}) and (\ref{tau}).$\Box$\\
\textbf{Remark 9}:\ So starting from the ``ghost" symmetry, we find
that $C\tau + \frac{1}{2}\sum_{a\in\{\alpha\}}
\Omega(\Phi_{2a}(t),\Phi_{1a}(t)) \tau(t)$ is a new $\tau$ function
of the BKP hierarchy. This transformation
is also given by Loris\cite{LorisWillox99} started from bilinear identity.\\
\textbf{Remark 10}:The symmetry reduction of BKP hierarchy,which is
now called constrained BKP(cBKP) hierarchy \cite{LorisWillox99} , is
just to identify $\pa_\alpha$ with $-\pa_{t_{2n+1}}$,i.e.
\begin{equation}\label{cBKPlax1}
    (L^{2n+1})_-=\sum_{a\in\{\alpha\}} \Phi_{2a}\pa^{-1}\Phi_{1a,x}-\Phi_{1a}\pa^{-1}\Phi_{2a,x},
\end{equation}
or
\begin{equation}\label{cBKPtau}
    \pa_{t_{2n+1}}\tau(t)=-
\frac{1}{2}\sum_{a\in\{\alpha\}}\Omega(\Phi_{2a}(t),\Phi_{1a}(t))\tau(t)
    =\frac{1}{2}\sum_{a\in\{\alpha\}}\Omega(\Phi_{1a}(t),\Phi_{2a}(t))\tau(t).
\end{equation}
Note if set $(L^{2n+1})_-=A$ as (\ref{cBKPlax1}), then
$\pa_{t_{2n+1}}L=[B_{2n+1},L]=[-L^{2n+1}_-, L]=-\pa_{\alpha}L$. So
$\pa_{\alpha}=-\pa_{t_{2n+1}}$.

To conclude this section, we would like to stress that the ``ghost "
symmetry (\ref{ghsym1}) of the BKP hierarchy is indeed different
from the counterpart in the KP hierarchy. This difference is due to
the BKP constraint (\ref{constr}). Moreover, the BSEP provides a
convenient tool to show it.

\section{Applications}
In this section, we shall show two applications for previous
results.

Firstly,let's derive a bilinear identity for the cBKP hierarchy
(\ref{cBKPlax1}) through the spectral representation of the BKP
hierarchy. Since
\begin{eqnarray}
  -\lambda^{2n+1} \psi_{BA}(t,-\lambda)&=& L^{2n+1}
  (\psi_{BA}(t,-\lambda))
  = (L^{2n+1})_+(\psi_{BA}(t,-\lambda))+(L^{2n+1})_-( \psi_{BA}(t,-\lambda))\nonumber\\
  &=&
   \pa_{t_{2n+1}}\psi_{BA}(t,-\lambda)+\pa_{\alpha}\psi_{BA}(t,-\lambda)\nonumber\\
  &=&\pa_{t_{2n+1}}\psi_{BA}(t,\lambda)+\frac{1}{2}
\sum_{a\in\{\alpha\}}[\Phi_{2a}(t)\Omega(\Phi_{1a}(t),\Psi_{BA}(t,-\lambda))\nonumber\\
&&-\Phi_{1a}(t)\Omega(\Phi_{2a}(t),\Psi_{BA}(t,-\lambda)) ] \ \
using\ (\ref{ghsym2}) \nonumber
\\
&=&\pa_{t_{2n+1}}\psi_{BA}(t,-\lambda)+\frac{1}{2}
\sum_{a\in\{\alpha\}}[-\Phi_{2a}(t)\Phi_{1a}(t+2[\lambda^{-1}])\nonumber\\
&& +\Phi_{1a}(t)\Phi_{2a}(t+2[\lambda^{-1}])]\Psi_{BA}(t,-\lambda).
~using~(\ref{Omegaphipsinegativelambda}) \label{bilinearproof}
\end{eqnarray}
So according to (\ref{bilinearproof}) and the bilinear identity of
the BKP hierarchy, we have
\begin{eqnarray}
&&\int
d\lambda\lambda^{2n}\psi_{BA}(t,\lambda)\psi_{BA}(t',-\lambda)\nonumber\\
&=&\sum_{a\in\{\alpha\}}\frac{1}{2}\int
d\lambda\lambda^{-1}[\psi_{BA}(t,\lambda)\psi_{BA}(t',-\lambda)\Phi_{2a}(t')\Phi_{1a}(t'+2[\lambda^{-1}])\nonumber\\
&&-\psi_{BA}(t,\lambda)\psi_{BA}(t',-\lambda)\Phi_{1a}(t')\Phi_{2a}(t'+2[\lambda^{-1}])]\nonumber\\
&=&\sum_{a\in\{\alpha\}}[\Phi_{2a}(t')\int
d\lambda\lambda^{-1}\psi_{BA}(t,\lambda)\psi_{BA}(t',-\lambda)[\frac{1}{2}\Phi_{1a}(t'+2[\lambda^{-1}])+\frac{1}{2}\Phi_{1a}(t')]-(1\leftrightarrow2)\nonumber\\
&=&\sum_{a\in\{\alpha\}}[\Phi_{2a}(t')\Phi_{1a}(t)-\Phi_{1a}(t')\Phi_{2a}(t)].\
\ \ \ using \ (\ref{srbkp3})\nonumber
\end{eqnarray}
Thus we get,
\begin{proposition}
For the constrained BKP hierarchies (\ref{cBKPlax1}),the bilinear
identity  can be written as
\begin{equation}\label{cBKPbilinear}
   \int
d\lambda\lambda^{2n}\psi_{BA}(t,\lambda)\psi_{BA}(t',-\lambda)=\sum_{a\in\{\alpha\}}[\Phi_{2a}(t')\Phi_{1a}(t)-\Phi_{1a}(t')\Phi_{2a}(t)].
\end{equation}
\end{proposition}
\noindent\textbf{Remark 11}: The bilinear identity  of the cBKP is
the same with Loris' paper\cite{LorisWillox99}.

Next,we will study the relation between the ``ghost" symmetry and
the additional symmetry. By using the ``ghost" symmetry of  the BKP
hierarchy, we shall give a simple proof of the
Adler-Shiota-van-Moerbeke formula\cite{L95,L96,Tu07} of the  BKP
which provides the connection between the form of additional
symmetries of the BKP hierarchy acting on BA functions and Sato
Backlund symmetry acting on the tau-functions of the BKP hierarchy.
To this end, let
$Y(\lambda,\mu)\equiv\psi_{BA}(t,-\lambda)\pa^{-1}\psi_{BA}(t,\mu)_x-\psi_{BA}(t,\mu)\pa^{-1}
\psi_{BA}(t,-\lambda)_x$ be pseudodifferential operator inducing a
special ``ghost" symmetry flow $\pa_{(\lambda,\mu)}W\equiv
Y(\lambda,\mu)W$ according to (\ref{ghsym1}). In this case,the
``ghost"  symmetry flow is generated by an infinite combination of
additional symmetries\cite{Tu07,Tu08}. Then,
$\pa_{(\lambda,\mu)}W\equiv Y(\lambda,\mu)W$ infers its actions on
wave function
$$\pa_{(\lambda,\mu)}(\psi_{BA}(t,z))= Y(\lambda,\mu)(\psi_{BA}(t,z)).$$
 Taking (\ref{ghsym2}) into it, we have
\begin{equation}\label{ghsym6}
Y(\lambda,\mu)(\psi_{BA}(t,z))=\frac{1}{2}\Big(\psi_{BA}(t,-\lambda)\Omega\big(\psi_{BA}(t,\mu),\psi_{BA}(t,z)\big)-
\psi_{BA}(t,\mu)\Omega\big(\psi_{BA}(t,-\lambda),\psi_{BA}(t,z)\big)\Big).
\end{equation}
Further, according to  (\ref{wavefun}) and (\ref{tau}), the action
of the vertex operator $\mathcal {X}(\lambda,\mu)$ on the BA
function $\psi_{BA}(t,z)$ is as follows

\begin{equation}\label{vopwf}
\mathcal
{X}(\lambda,\mu)\psi_{BA}(t,z)=\psi_{BA}(t,z)\Delta_z\frac{\mathcal
{X}(\lambda,\mu)\tau(t)}{\tau(t)}.
\end{equation}
Now, the above results allow us to establish the connection between
$\mathcal{X}$ and $Y$.
\begin{proposition}
\begin{equation}\label{bkpasvm}
\mathcal{X}(\lambda,\mu)\psi_{BA}(t,z)=2Y(\lambda,\mu)\psi_{BA}(t,z).
\end{equation}
\end{proposition}
\noindent{\bf Proof}:
\begin{eqnarray*}
        &&\mathcal{X}(\lambda,\mu)\psi_{BA}(t,z)= \psi_{BA}(t,z)\Delta_z\frac{\mathcal
{X}(\lambda,\mu)\tau(t)}{\tau(t)} \\
        &=& - \psi_{BA}(t,z)  \psi_{BA}(t,-\lambda) \psi_{BA}(t-2[z^{-1}],\mu)+ \psi_{BA}(t,z)  \psi_{BA}(t,\mu)
        \psi_{BA}(t-2[z^{-1}],-\lambda)\\
        && using\ (\ref{ba2})\ and \ (\ref{shdo})\\
        &=&\psi_{BA}(t,-\lambda)\Omega(
        \psi_{BA}(t,\mu),\psi_{BA}(t,z))-\psi_{BA}(t,\mu)\Omega(
        \psi_{BA}(t,-\lambda),\psi_{BA}(t,z))\\
        && \ using \  (\ref{bkpsep1})\\
        &=&2Y(\lambda,\mu)(\psi_{BA}(t,z)).~ using~(\ref{ghsym6}) \mbox{\hspace{3cm}} \Box
      \end{eqnarray*}
\appendix
\section{ Proof of (\ref{corollary3}) }\label{appcorollary3}
\begin{eqnarray*}
  S(\psi_{BA}(t,\lambda),\Phi_x(t)) &=& S(\psi_{BA}(t,\lambda),\int
  d\mu\mu^{-1}\varphi(\mu)\psi_{BA}(t,\mu)_x)~using\ \ (\ref{srbkp1})\\
  &=&\int d\mu\mu^{-1}\varphi(\mu)S(\psi_{BA}(t,\lambda),\psi_{BA}(t,\mu)_x)\\
  &=&\int
  d\mu\mu^{-1}\varphi(\mu)\frac{1}{2}(\psi_{BA}(t,\mu)-\psi_{BA}(t-2[\lambda^{-1}],\mu))\psi_{BA}(t,\lambda)~using\ \ (\ref{corollary1})\\
    &=&\frac{1}{2}\psi_{BA}(t,\lambda)\big(\Phi(t)-\Phi(t-2[\lambda^{-1}])\big).\ \  using\ (\ref{srbkp1})\\
\end{eqnarray*}
\section{ Proof of (\ref{corollary4}) }\label{appcorollary4}
\begin{eqnarray*}
S(\Phi_1(t),\Phi_{2x}(t)) &=&S(\int
d\mu\mu^{-1}\varphi_1(\mu)\psi_{BA}(t,\mu),\int d\lambda\lambda^{-1}
   \varphi_2(\lambda)\psi_{BA}(t,\lambda)_x)~using\ \ (\ref{srbkp1})\\
 &=&\int\int d\lambda d\mu \lambda^{-1}\mu^{-1}
  \varphi_1(\mu) \varphi_2(\lambda)S\big(\psi_{BA}(t,\mu),\psi_{BA}(t,\lambda)_x\big).
\end{eqnarray*}
\section{ Proof of (\ref{flowcommutativity2}) }\label{appflowcommutativity2}
According to (\ref{zhengming3}),
\begin{eqnarray*}
  &&Res(AB_{2n+1}\Phi\pa^{-1}) \\
  &=&Res(\pa_\alpha B_{2n+1}\Phi\pa^{-1})-Res(\pa_{t_{2n+1}} A\Phi\pa^{-1})+Res(B_{2n+1}A\Phi\pa^{-1}) \\
   &=&(\pa_\alpha B_{2n+1})(\Phi)+Res(B_{2n+1}A\Phi\pa^{-1})~~~using~(\ref{usefulformua3})\\
   &=&(\pa_\alpha B_{2n+1})(\Phi)+\sum\limits_{a\in\{\alpha\}}Res(B_{2n+1}(\Phi_{2a}\pa^{-1}\Phi_{1a,x}-\Phi_{1a}\pa^{-1}\Phi_{2a,x})\Phi\pa^{-1})\\
   &=&(\pa_\alpha B_{2n+1})(\Phi)+\sum\limits_{a\in\{\alpha\}}\Big(Res(B_{2n+1}\Phi_{2a}\pa^{-1}(\pa
   S(\Phi,\Phi_{1a,x})\\
   &&-S(\Phi,\Phi_{1a,x})\pa)\pa^{-1})-(1\leftrightarrow
   2)\Big)~~~using~(\ref{usefulformua2})\\
   &=&(\pa_\alpha B_{2n+1})(\Phi)+\sum\limits_{a\in\{\alpha\}}\Big(Res(B_{2n+1}\Phi_{2a}S(\Phi,\Phi_{1a,x})\pa^{-1})\\
   &&-Res(B_{2n+1}\Phi_{2a}\pa^{-1}S(\Phi,\Phi_{1a,x}))-(1\leftrightarrow
   2)\Big)\\
   &=&(\pa_\alpha B_{2n+1})(\Phi)+\sum\limits_{a\in\{\alpha\}}\Big(B_{2n+1}(\Phi_{2a}S(\Phi,\Phi_{1a,x}))\\
   &&-B_{2n+1}(\Phi_{2a})S(\Phi,\Phi_{1a,x})-(1\leftrightarrow
   2)\Big)~using ~(\ref{usefulformua3})\\
  &=&(\pa_\alpha B_{2n+1})(\Phi)+B_{2n+1}\sum\limits_{a\in\{\alpha\}}\Big(\Phi_{2a}S(\Phi,\Phi_{1a,x})-\Phi_{1a}S(\Phi,\Phi_{2a,x})\Big)\\
  &&-\sum\limits_{a\in\{\alpha\}}\Big((\pa_{t_{2n+1}}\Phi_{2a})S(\Phi,\Phi_{1a,x})-(\pa_{t_{2n+1}}\Phi_{1a})S(\Phi,\Phi_{2a,x})\Big)\\
  &=&(\pa_\alpha B_{2n+1})(\Phi)+B_{2n+1}(\pa_\alpha\Phi)
  -\sum\limits_{a\in\{\alpha\}}\Big((\pa_{t_{2n+1}}\Phi_{2a})S(\Phi,\Phi_{1a,x})-(\pa_{t_{2n+1}}\Phi_{1a})S(\Phi,\Phi_{2a,x})\Big)\\
  &&~~~~~using~(\ref{proof})\\
&=&\pa_\alpha(\pa_{t_{2n+1}}\Phi)-\sum\limits_{a\in\{\alpha\}}\Big((\pa_{t_{2n+1}}\Phi_{2a})S(\Phi,\Phi_{1a,x})
-(\pa_{t_{2n+1}}\Phi_{1a})S(\Phi,\Phi_{2a,x})\Big).
\end{eqnarray*}
{\bf {Acknowledgements:}}
  {\small   This work is supported by the NSFC(10671187, 10971109) and Program for NCET under Grant
  No.NCET-08-0515. Sen Hu is partially supported by NSFC (10771203)and a renovation project from the Chinese
  Academy of Sciences  We thank Professor Yishen Li  and Yi Cheng(USTC,China) for long-term encouragements
  and supports. We also thank Dr.Chen Chunli(SJTU,China) for her suggestions. We also thank anonymous referee for his/her
  useful comments.}

\end{document}